\documentclass[aps,prl,reprint,groupedaddress]{revtex4-1}
\usepackage{mathtools}
\usepackage{amsmath}
\usepackage{amssymb}
\usepackage{color}
\begin{document}

\title{Restricted Euler dynamics along trajectories of small inertial particles in turbulence}


\author{Perry L. Johnson}
\email{pjohns86@jhu.edu}
%\thanks{}
%\affiliation{Department of Mechanical Engineering, Johns Hopkins University}

\author{Charles Meneveau}
%\thanks{}
\affiliation{Department of Mechanical Engineering and Center for Environmental and Applied Fluid Mechanics, The Johns Hopkins University, 3400 N. Charles Street, Baltimore, Maryland 21218, USA}


\date{\today}

\begin{abstract}

The fate of small particles in turbulent flows depends strongly on the surrounding fluid's velocity gradient properties such as rotation and strain-rates.  For non-inertial (fluid) particles, the Restricted Euler model provides a simple, low-dimensional dynamical system representation of Lagrangian evolution of velocity gradients in fluid turbulence, at least for short times. Here we derive a new restricted Euler dynamical system for the velocity gradient evolution of inertial particles such as solid particles in a gas or droplets and bubbles in turbulent liquid flows.  The model is derived in the limit of small (sub Kolmogorov scale) particles and low Stokes number. The system exhibits interesting fixed points, stability and invariant properties. Comparisons with data from Direct Numerical Simulations show that the model predicts realistic trends such as the tendency of increased straining over rotation along heavy particle trajectories and, for light particles such as bubbles, the tendency of reduced self-stretching of strain-rate.

\end{abstract}

\pacs{}
%\keywords{}

\maketitle

%\section{Introduction}

Small particles embedded in a turbulent flow have interesting behaviors when the particle density, $\rho_p$, is different from the density of the surrounding fluid, $\rho_f$. For example, within a certain parameter range, heavy particles tend to cluster in regions where the strain-rate is higher than the rotation-rate \cite{Maxey1987,Wang1993,Eaton1994,Monchaux2012}, while the opposite is true of lighter particles \cite{Biferale2010} such as bubbles \cite{Wang1993b,Calzavarini2008} and oil droplets \cite{Gopalan2008}. This effect can drive heavy particles to cluster toward fractal sets \cite{Balkovsky2001,Bec2003}, which can enhance collision rates \cite{Sundaram1997,Reade2000,Wang2000,Falkovich2002,Bewley2013}. The rate of fractal clustering can be related to the surrounding fluid's velocity gradient structure experienced by particles along their trajectories \cite{Maxey1987,Balkovsky2001}. Other important aspects of multi-phase flows in various applications, such as particle rotation and orientation \cite{Pumir2011,Parsa2012,Chevillard2013}, droplet or bubble deformation \cite{Maffettone1998,Biferale2014}, and nutrient uptake \cite{Batchelor1980,Karp-Boss1996} similarly depend on the local velocity gradient structure. Much of recent research on particle evolution in fluid turbulence \cite{Toschi2009} is based on direct numerical simulations (DNS). However, the high-dimensionality of the Navier-Stokes equations especially for high Reynolds number turbulence \cite{Ishihara2009,Ireland2016} complicates basic analysis and the development of physical insights. 
Dynamical systems models for the velocity gradient along Lagrangian paths provide an interesting possibility for reducing turbulent dynamics to a low-dimensional representation. Vieillefosse \cite{Vieillefosse1982,Vieillefosse1984} and Cantwell \cite{Cantwell1992} developed and studied the so-called restricted Euler system, which is obtained by taking the spatial gradient of the Navier Stokes equations and neglecting the viscous and anisotropic pressure Hessian contributions. The model consists of a system of $3\times 3$ nonlinear coupled ordinary differential equations for velocity gradient tensor elements 
\begin{equation}
\frac{DA_{ij}}{Dt} = -A_{ik}A_{kj} + \frac{1}{3} A_{k\ell} A_{\ell k} \delta_{ij}, ~~~ i,j = 1,2,3,
\end{equation}
where $A_{ij} = \partial u_i / \partial x_j$ is the fluid velocity gradient, $u_i({\bf x},t)$ is the velocity field, and $D/Dt = \partial/\partial t + u_k \partial/\partial x_k$ represents the Lagrangian time derivative following a fluid element in the flow.
%LONG: This rather simple system can be further projected onto two-dimensional invariant space,
%\begin{equation}
%\frac{DQ}{dt} = -3R, \hspace{0.1\linewidth} \frac{DR}{dt} = \frac{2}{3}Q^2,
%\end{equation}
%where $Q = - \tfrac{1}{2} A_{ij} A_{ji}$ and $R = - \tfrac{1}{3} A_{ij}A_{jk}A_{ki}$ are invariants of the velocity gradient tensor.

The restricted Euler system was shown to display important features seen in turbulent flows, such as the preferential alignment of the vorticity vector in the direction of the eigenvector associated with the median eigenvalue of the strain-rate \cite{Kerr1985,Ashurst1987}, negative skewness in longitudinal velocity gradients, as well as the tendency to produce extreme velocity gradient events \cite{Meneveau2011}, which are clustered along the so-called Vieillefosse tail in the so-called RQ-invariant phase-space. Without the neglected, unclosed  terms, however, the restricted Euler system eventually yields finite time singularities for almost all initial conditions. Subsequent work  on modeling the unclosed terms  \cite{Girimaji1990,Jeong2003,Chevillard2006,Biferale2007,Chevillard2008,Wilczek2014,Johnson2016b} and related work on the perceived velocity gradient \cite{Chertkov1999,Pumir2013} at various scales  
has resulted in models capable of reproducing certain turbulent statistics with some accuracy, although extension to arbitrarily high Reynolds numbers remains an open challenge \cite{MartinsAfonso2010,Meneveau2011}.

Using DNS, Benzi et al. \cite{Benzi2009} studied empirically the impact of particle inertia on the fluid velocity gradient probability density in the subspace formed by the two tensor invariants $Q=-\frac{1}{2}{\rm Tr}({\bf A}^2)$ and 
$R=-\frac{1}{3}{\rm Tr}({\bf A}^3)$.  For small particles much lighter than the surrounding fluid (e.g. small bubbles), the tendency of particles to visit velocity gradients along the Vieillefosse tail is dramatically reduced, and $\langle Q \rangle > 0$. The opposite is true for particles much heavier than the surrounding fluid that tend to experience higher probabilities for more extreme states along the Vieillefosse tail and $\langle Q \rangle < 0$. In this Letter, we derive an extension to the restricted Euler system that considers the effect of inertia on the velocity gradient dynamics when following an inertial particle and explore whether the trends observed in Ref. \cite{Benzi2009} can be explained by the proposed  low-dimensional model. 
 
% \cite{Ireland2016,Ireland2016a}.
%LONG: Like the original restricted Euler, the resulting dynamical system can be further projected onto two-dimensional invariant space. As such, the analysis becomes straight-forward and the intuitive understanding of such a system is enhanced. This extension of the restricted Euler system provides a promising starting point for the modeling of velocity gradients along inertial trajectories.

%\section{Model}

%LONG: Consider the trajectory, $y_i(t)$, of a particle with radius $a$ and density $\rho_p$ embedded in a turbulent flow with fluid density $\rho_f$, viscosity $\mu = \rho_f \nu$, and mean dissipation rate $\langle \epsilon \rangle$. The turbulent flow is fully described by its velocity field, $u_i(\mathbf{x},t)$, which is a solution to the incompressible Navier-Stokes equations, $Du_i/Dt = -\partial p/\partial x_i + \nu \nabla^2 u_i + g_i + f_i$, $\partial u_k/\partial x_k = 0$, with constant gravitational body force $g_i$, arbitrary forcing $f_i(\mathbf{x},t)$, and appropriate boundary and initial data. The pressure field, $p(\mathbf{x},t)$, is given as the non-local integral solution to the pressure Poisson equation, $\nabla^2 p = - \partial u_i/\partial x_j ~\partial u_j/\partial x_i$.

\begin{figure}[htbp]
\begin{center}
\includegraphics[width=0.7\linewidth]{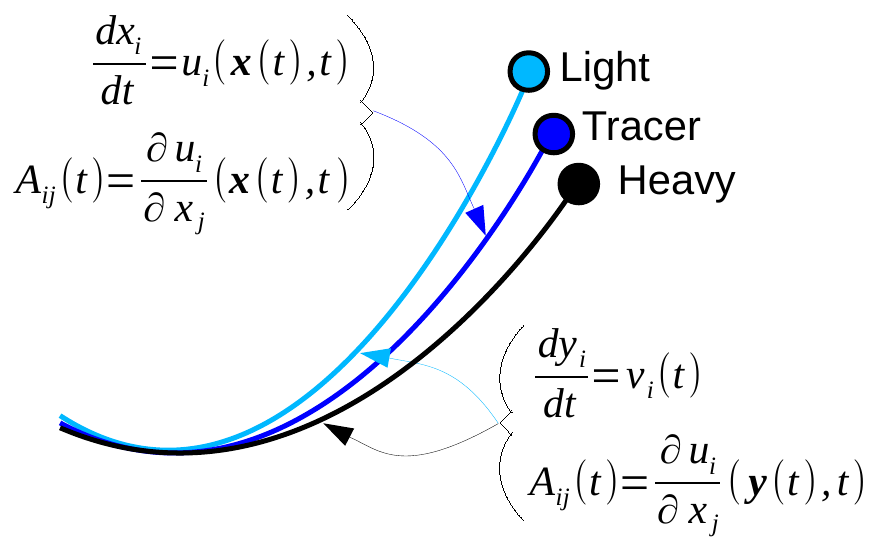}
\caption{Sketch of fluid and inertial particle trajectories. In this Letter, we consider the time history of fluid velocity gradients, $A_{ij}(t)$, along these trajectories.}
\label{fig:trajectory_artwork}
\end{center}
\end{figure}

As illustrated in Figure \ref{fig:trajectory_artwork}, while fluid tracers (position ${\bf x}(t)$) move according to $dx_i/dt = u_i(\mathbf{x},t)$, inertial particle trajectories (${\bf y}(t)$) evolve following the particle velocity ${\bf v}(t)$ according to $dy_i/dt = v_i(t)$, where in general, $v_i(t) \neq u_i(\mathbf{y},t)$. When the particle radius $a \ll \eta = \nu^{3/4}\langle\epsilon\rangle^{-1/4}$ (Kolmogorov scaled, where $\nu$ is kinematic viscosity and $\epsilon$ is the dissipation rate) and $Re_a = a |\mathbf{v}-\mathbf{u}|/\nu \ll 1$ (particle Reynolds number),
%the drag force can be represented using Stokes formula, $F_i = 6\pi\mu a(u_i-v_i)$, and
the dynamical equation of the inertial particle trajectory \cite{Maxey1983} can be simplified to \cite{Maxey1987,Balkovsky2001},
\begin{equation}
\frac{dv_i}{dt} = \beta \frac{Du_i}{Dt} + \frac{u_i-v_i+v_i^\infty}{\tau_p},
\end{equation}
where $\beta = 3\rho_f/(2\rho_p + \rho_f)$ is the added mass parameter, $\tau_p = a^2/3\nu\beta$ is the relaxation time for the trajectory of a spherical particle of radius $a$, and $v_i^{\infty} = (1-\beta) \tau_p g_i$ is the terminal velocity. For small Stokes number based on the Kolmogorov timescale ($\tau_\eta = \nu^{1/2}\langle\epsilon\rangle^{-1/2}$), $St = \tau_p/\tau_\eta \ll 1$, a perturbation solution yields the following approximation \cite{Maxey1987,Balkovsky2001},
\begin{equation}
v_i = u_i + v_i^\infty - (1-\beta) \tau_p \frac{Du_i}{Dt} - \tau_p v_j^\infty \frac{\partial u_i}{\partial x_j}.
\label{eq:perturb_sol}
\end{equation}
This solution admits an interpretation in terms of a particle velocity field, $v_i(\mathbf{x},t)$, such that the velocity of a particle at location $y_i(t)$ can be approximated by $v_i(t) = v_i(\mathbf{y}(t),t)$. In this way, the particle's time derivative can be interpreted as $d/dt = \partial/\partial t + v_k \partial/\partial x_k$.

In this Letter, we consider the evolution of the fluid velocity gradient, $A_{ij} = \partial u_i/\partial x_j$, along the particle trajectory, as sketched in Figure \ref{fig:trajectory_artwork}. Considering a particle velocity field $v_i(\mathbf{x},t)$, the evolution equation for the velocity gradient can be related to the Lagrangian evolution by $dA_{ij}/dt = DA_{ij}/Dt + \left( v_k - u_k \right) \partial A_{ij}/\partial x_k$, which upon substitution of the gradient of Navier-Stokes yields,
%\begin{align}
%\frac{dA_{ij}}{dt} = -A_{ik}A_{kj}-\frac{\partial^2 p}{\partial x_i \partial x_j} + \nu \nabla^2 A_{ij} + \frac{\partial f_i}{\partial x_j} \nonumber\\
%+ \frac{\partial}{\partial x_k} \left[ \left( v_k - u_k \right) A_{ij} \right] - \frac{\partial v_k}{\partial x_k} A_{ij}.
%\end{align}
\begin{equation}
\frac{dA_{ij}}{dt} = -A_{ik}A_{kj}-\frac{\partial^2 p}{\partial x_i \partial x_j} -  \frac{\partial v_k}{\partial x_k} A_{ij}  - \frac{\partial T_{ijk}}{\partial x_k},
\end{equation}
where $p$ is the pressure divided by density and $T_{ijk}$ represents spatial fluxes of velocity gradient due to viscosity, terminal velocity and inertial effects according to $T_{ijk} = - \nu \partial A_{ij}/\partial x_k - A_{ij} \left[v_k^\infty - (1-\beta) \tau_p Du_k/Dt - \tau_p v_\ell^\infty A_{k\ell} \right]$. 
A key step is to evaluate the particle velocity-field divergence (i.e. divergence of Eq.  \eqref{eq:perturb_sol}) for a divergence-free fluid velocity field \cite{Balkovsky2001}, i.e.  $\partial v_k/\partial x_k = (1-\beta) \tau_p A_{k \ell}A_{\ell k}$.  The final steps in deriving the new inertial restricted Euler system are, similarly as in the classical Restricted Euler model, (a) to replace the pressure Hessian $\partial_i\partial_j p$ by its isotropic part $\nabla^2 p \, (\delta_{ij}/3)$ and to invoke the pressure Poisson equation $\nabla^2 p = - A_{k \ell}A_{\ell k}$, and (b) to neglect any spatial fluxes, i.e. setting $T_{ijk}=0$ where we make the strong assumption of neglecting both fluxes due to viscosity as well as due to terminal particle velocity and other inertia effects. 

%Substitution of \eqref{eq:perturb_sol} for the particle velocity field yields,
%\begin{align}
%\frac{dA_{ij}}{dt} = -A_{ik}A_{kj}-\frac{\partial^2 p}{\partial x_i \partial x_j} + \nu \nabla^2 A_{ij} + \frac{\partial f_i}{\partial x_j}\nonumber\\
%+ \frac{\partial T_{ijk}}{\partial x_k} + (1-\beta) \tau_p A_{k\ell} A_{\ell k} A_{ij}.
%\end{align}
%where $T_{ijk} = A_{ij} \left(v_k^\infty - (1-\beta) \tau_p Du_k/Dt - \tau_p v_\ell^\infty A_{k\ell} \right) $.
%This equation represents an attractive low-dimensional representation of certain features in particle-laden turbulence, but it contains unclosed that are difficult to model. Various closure models for the pressure Hessian and viscous Laplacian terms have been proposed in the context of Lagrangian velocity gradient evolution \cite{Meneveau2011}, but now an additional unclosed term has been introduced. However, the last term on the right-hand side is closed and represents part of the inertial effects. Following Vieillefosse \cite{Vieillefosse1982,Vieillefosse1984} and Cantwell \cite{Cantwell1992}, a simple model respecting incompressibility can be constructed by assuming an isotropic pressure Hessian, $\partial^2 p/\partial x_i \partial x_j = - \frac{1}{3} A_{k\ell}A_{\ell k} \delta_{ij}$, and neglecting the other divergence terms, $\nu \nabla^2 A_{ij} + \partial f_i/\partial x_j + \partial T_{ijk}/\partial x_k = 0$. 

The resulting system reads as follows,
\begin{equation}
\frac{dA_{ij}}{dt} = -A_{ik}A_{kj} + \frac{1}{3} A_{k\ell}A_{\ell k} \delta_{ij} + (1-\beta) \tau_p A_{k\ell} A_{\ell k} A_{ij},
\label{eq:main}
\end{equation}
thus extending the restricted Euler system of equations to include inertial trajectory effects. The original restricted Euler equation is recovered by considering particles with equal density to the surrounding fluid, $\rho_p = \rho_f$, hence $\beta = 1$.

\begin{figure}[htbp]
\begin{center}
\includegraphics[width=0.8\linewidth]{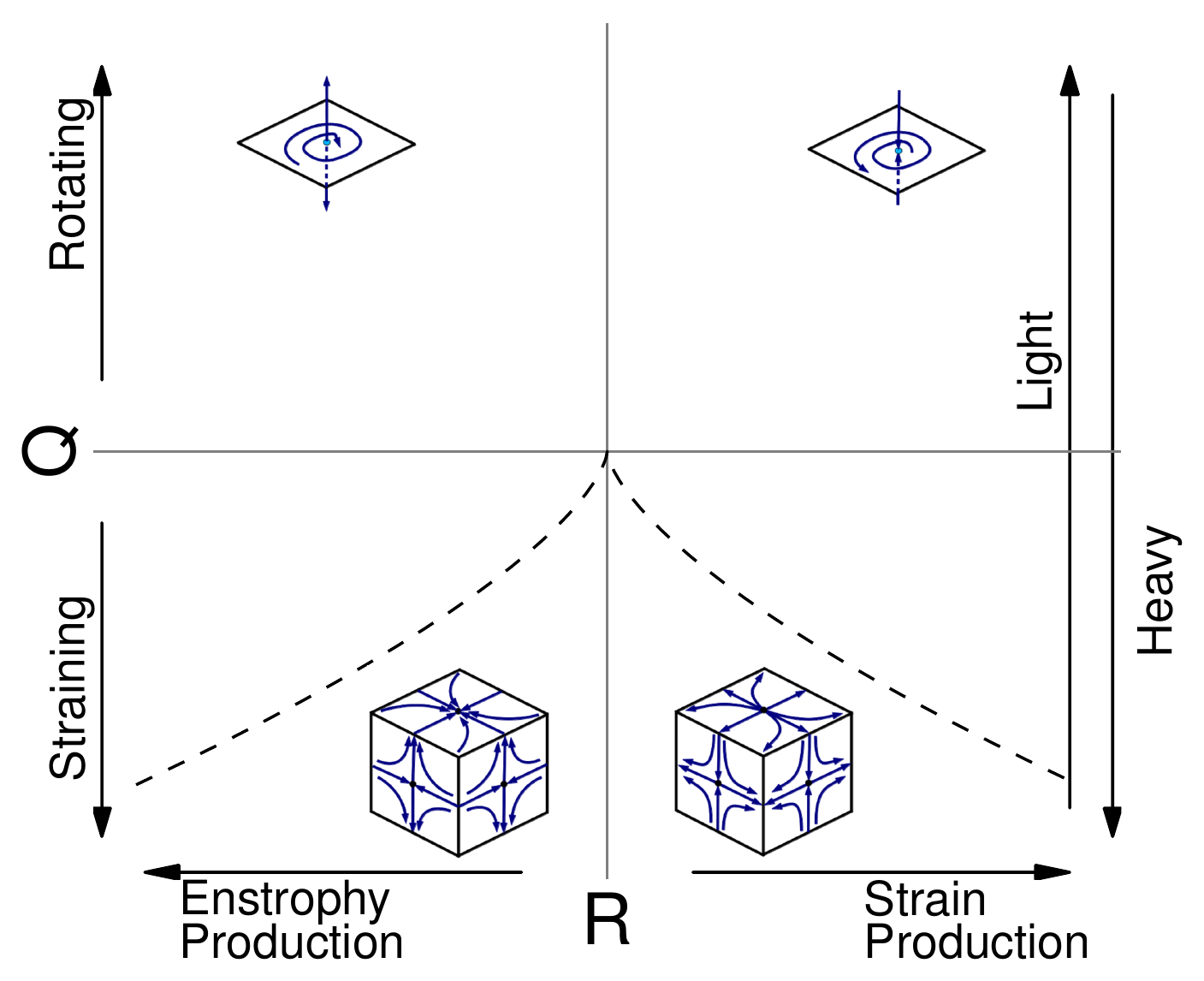}
\caption{Sketch outlining the features of the $RQ$ invariant space, including representative local flow topology cubes. The Vieillefosse tail (dashed line) represents the boundary between real and complex eigenvalues of the velocity gradient tensor.}
\label{fig:QR_artwork}
\end{center}
\end{figure}

The inertial  restricted Euler dynamics given by \eqref{eq:main} can be projected into the two-dimensional space of tensor invariants $Q$ and $R$, and yields the following two-dimensional dynamical system:
\begin{equation}
\frac{dQ}{dt} = -3R - \frac{2}{3} \alpha Q^2, \hspace{0.1\linewidth} \frac{dR}{dt} = \frac{2}{3}Q^2 - \alpha Q R,
\label{eq:QR}
\end{equation}
where $\alpha = 6 (1-\beta) \tau_p$ is the timescale representing inertial effects. The second invariant, $Q = \tfrac{1}{2} \left( \Omega_{ij}\Omega_{ij} - S_{ij}S_{ij} \right)$, represents the relative balance between local rotation, $\Omega_{ij} = \tfrac{1}{2} \left( A_{ij} - A_{ji} \right)$, and straining, $S_{ij} = \tfrac{1}{2}\left( A_{ij} + A_{ji} \right)$. The third invariant, $R = -\tfrac{1}{3} S_{ij}S_{jk}S_{ki} - \tfrac{1}{4}\omega_{i}S_{ij}\omega_{j}$, represents the competition of strain production and enstrophy production \cite{Chevillard2008}. 
For particles that are heavier than the surrounding fluid, $0 < \beta < 1$ and $\alpha > 0$. For particles lighter than the surrounding fluid, $1 < \beta < 3$ and $\alpha < 0$.  For heavy particles ($\alpha > 0$), the inertial term in the evolution equation for $Q$ tends to oppose rotation-dominant states ($Q > 0$) and reinforce strain-dominant states ($Q < 0$). The exact opposite is true for light particles, where the inertial term opposes highly straining states and favors highly rotating states. In this way, heavy particles cluster in straining regions ($Q < 0$) and lighter particles cluster in rotating regions ($Q > 0$), qualitatively mimicking well-known preferential concentration trends. The qualitative features of $RQ$ space including the effects of inertia are sketched in Figure \ref{fig:QR_artwork}.  In homogeneous turbulence, $\langle Q \rangle = 0$ and $\langle R \rangle = 0$, where angle brackets denote ensemble averaging \cite{Betchov1956}. The standard ensemble averaging can be represented by averaging over fluid (non-inertial)  particle Lagrangian trajectories, but not for the case of inertial particle trajectories. When averaging over inertial trajectory ensembles, one observes that $\langle Q \rangle < 0$ for heavy particles and $\langle Q \rangle > 0$ for light particles \cite{Ireland2016}.

%\section{Results}

\begin{figure}[htbp]
\begin{center}
%Tracer \hspace{0.25\linewidth} Heavy \hspace{0.25\linewidth} Light\\
%\includegraphics[width=0.32\linewidth]{images/streamplot_restrictedEuler.eps}
%\includegraphics[width=0.32\linewidth]{images/streamplot_positiveAlpha.eps}
%\includegraphics[width=0.32\linewidth]{images/streamplot_negativeAlpha.eps}\\
%\includegraphics[trim = 112mm 224mm 21mm 26mm, clip, width=0.98\linewidth]{images/Benzi2009.pdf}\\
%\caption{Restricted Euler streamlines (top row) and DNS-computed joint-PDFs (bottom row) in $RQ$ invariant space for Lagrangian trajectories (left column), heavy particle trajectories (center column), and light particle trajectories (bottom column). In the streamline plots, the red circles represent fixed points of the restricted Euler dynamics and the timescale $|\alpha| = 6 |1-\beta| \tau_p$ is used to normalize the axes. DNS results are reproduced {\color{blue} with permission} from Benzi et al. \cite{Benzi2009}. For the DNS results, a Stokes number of $St = \tfrac{\tau_p}{\tau_\eta} = 0.5$ is used for heavy ($\beta = 0$) and light ($\beta = 3$) particles. The DNS results use the timescale $\langle Q^2 \rangle^{-1/4}$ to normalize the axes.}
\includegraphics[width=0.50\linewidth]{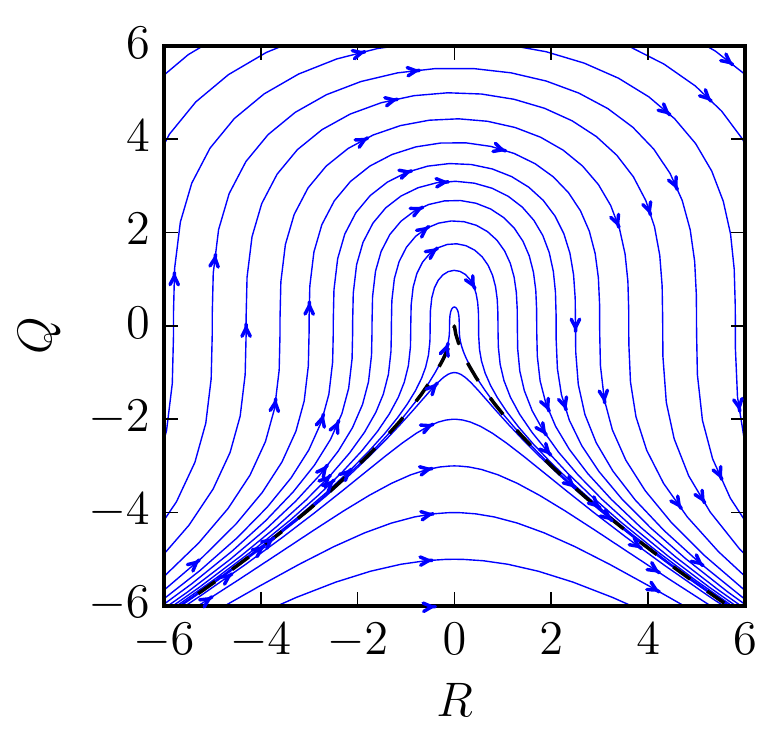}
\includegraphics[width=0.48\linewidth]{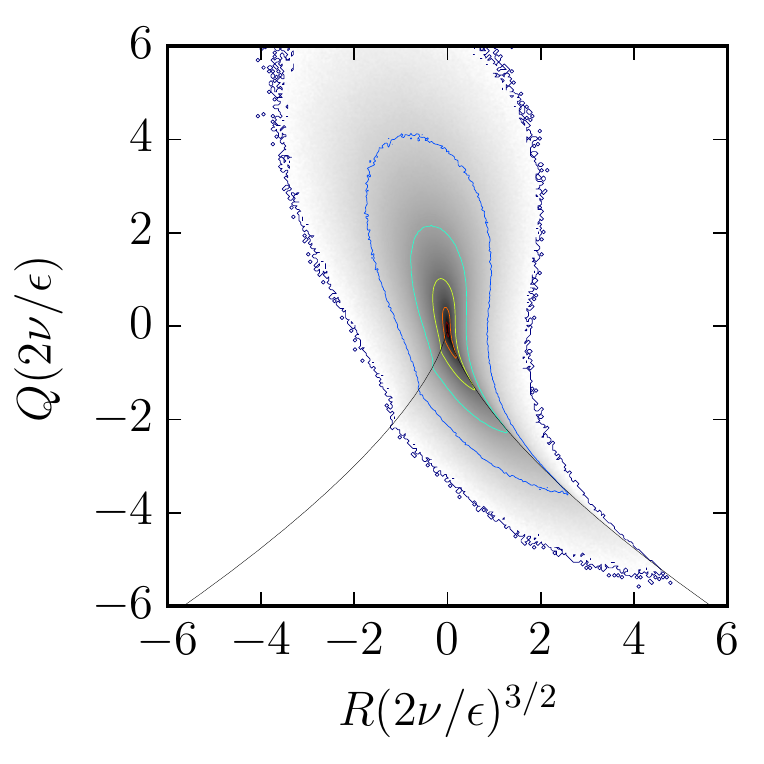}\\
\includegraphics[width=0.50\linewidth]{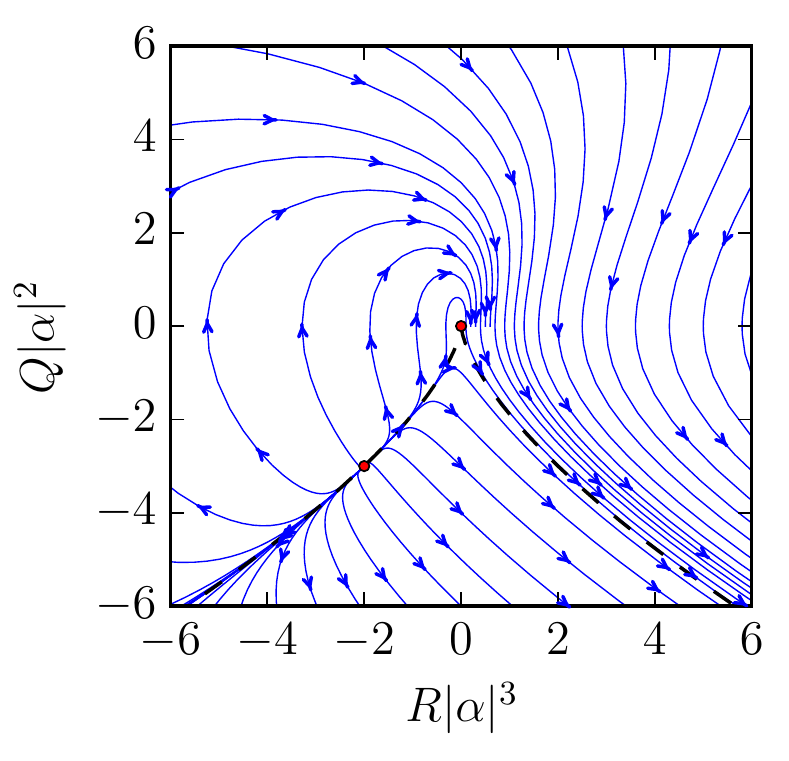}
\includegraphics[width=0.48\linewidth]{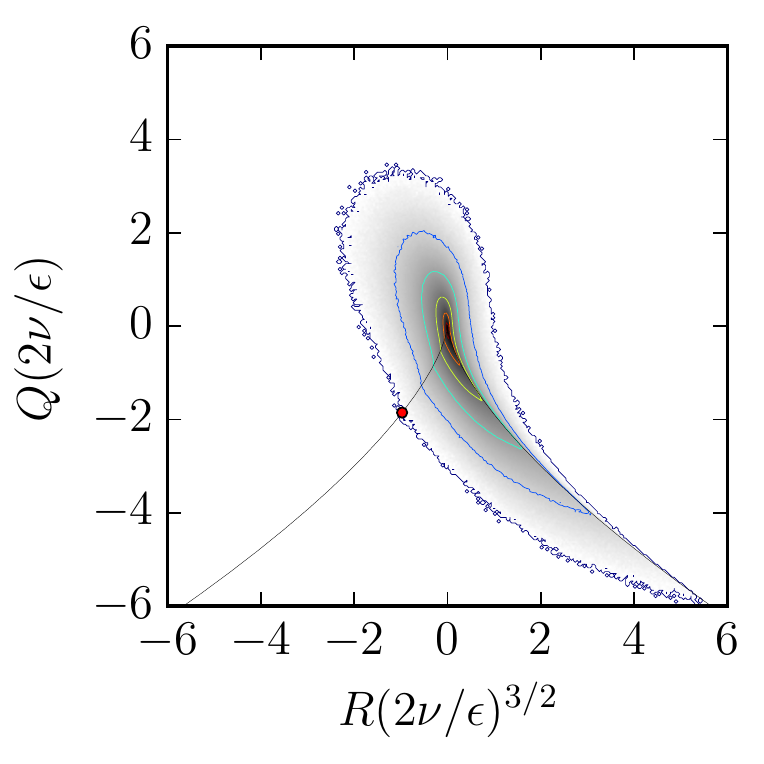}\\
\includegraphics[width=0.50\linewidth]{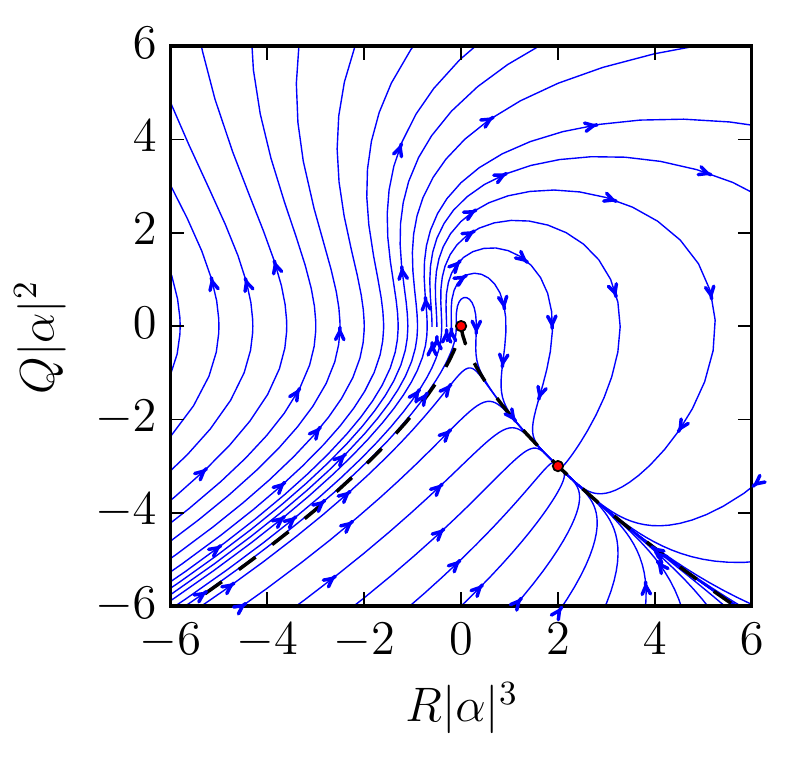}
\includegraphics[width=0.48\linewidth]{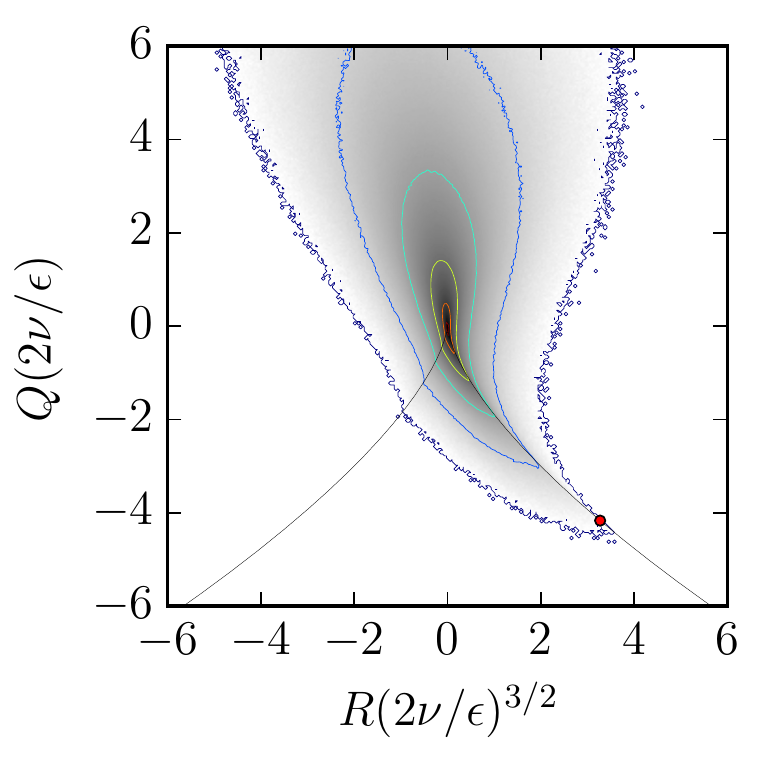}\\
\caption{Restricted Euler streamlines (left) and DNS-computed joint-PDF iso-contours (right) for Lagrangian trajectories (top), heavy particle trajectories with $\beta = 0$, $St = 0.3, \alpha = 1.8 \sqrt{\nu/\epsilon}$ (middle), and light particle trajectories with $\beta = 3$, $St = 0.1, \alpha = -1.2 \sqrt{\nu/\epsilon}$ (bottom).  The timescale $|\alpha|$ is used to normalize the axes on the streamline plots, while $\sqrt{2\nu / \epsilon}$ is used to normalize the axes for the DNS results, where $\epsilon$ is the turbulent dissipation rate from the simulation. The red circles show fixed points of the RQ dynamics, providing a visual connection between the two normalizations. The DNS data is from a pseudo-spectral simulation performed at $Re_\lambda = 185$ with a grid resolution of $512^3$ \cite{Bec2010}. The PDF iso-contours are spaced logarithmically with levels $10^{z}$, $z = 1, 0, -1, -2, -3, -4$.}
\label{fig:QRstreamplot}
\end{center}
\end{figure}

Figure \ref{fig:QRstreamplot} shows the RQ phase-space portrait for non-inertial (fluid tracer), heavy, and light particles computed numerically from \eqref{eq:QR}. 
% using a fourth-order Dormond-Price integration scheme with automatically adjusting time step size \cite{Hairer1993}, 
Also shown is the stationary joint-PDF of $Q$ and $R$ computed from DNS at $Re_\lambda = 185$ \cite{Bec2010}. Although statistical stationarity (and hence direct comparison of the joint-PDF) cannot be achieved in the system of Eq. \ref{eq:QR} without introducing models for the neglected terms, the qualitative comparison of streamlines with the joint-PDF in RQ space for heavy particles from DNS is informative. In particular, on the top left is the original restricted Euler system ($\alpha = 0$), for which trajectories move left to right along lines of constant $Q^3 + \frac{27}{4}R^2$, eventually proceeding toward the finite-time singularity in the fourth quadrant \cite{Vieillefosse1982,Vieillefosse1984,Cantwell1992}.  The sheared tear-drop shape in the joint-PDF on the top right highlights the dynamical significance of the Vieillefosse tail for the full dynamics of the velocity gradient tensor \cite{Soria1994,Blackburn1996,Chong1998,Ooi1999}.

In the middle row of Figure \ref{fig:QRstreamplot}, the inertial restricted Euler phase-space portrait is shown for the case of heavy particles ($\alpha > 0$). The finite-time singularity down the Vieillefosse line in the fourth quadrant remains and is strengthened. In addition, a new singularity is introduced down the other branch of the Vieillefosse line in the third quadrant, however it is a very unstable manifold in the third quadrant, meaning that any noise in the system will prevent particles from proceeding to that singularity. In the first quadrant, the downward ``flow'' of particles is enhanced while the left-to-right ``flow'' is suppressed. The DNS results for heavy particles indeed show the tendency down the Vieillefosse tail in the fourth quadrant, as well as reduced probabilities in the upper half ($Q > 0$).

Finally, the phase-space  trajectories for light particles ($\alpha < 0$) are shown on the bottom row of Figure \ref{fig:QRstreamplot}. The restricted Euler trajectories tend to proceed toward the fixed point in the fourth quadrant. There, a rapid collapse towards Vieillefosse tail is followed by slower evolution along it towards the fixed point. The restricted Euler dynamics impose more resistance to (e.g. noise-driven) movement away the tail than movement along the tail. The finite-time singularity down the Vieillefosse tail is regularized. However, some trajectories far enough to the left of the Vieillefosse tail in the second and third quadrants (e.g. $R(0)|\alpha|^3 < -3.2$ with $Q(0)|\alpha|^2 = 0$) do not circle around to the fixed point, but rather proceed to a finite-time singularity with $Q > 0$, asymptotically following inverted Vieillefosse-like manifolds with $Q \sim R^{2/3}$. The joint-PDF from DNS data indeed suggests that the Viellefosse tail is still dynamically important for light particles, but that light particles do not tend to reach extreme states as far down the Viellefosse tail compared with neutral and heavy particles, an effect that may be qualitatively linked to the fixed point in the restricted Euler dynamics. In general, the lower probabilities in the $Q<0$ region are offset by higher probabilities in the $Q>0$ region. Additionally, the upward and left-to-right movement in the first quadrant (toward $R \gg 0$) of the inertial restricted Euler streamlines is consistent with the enhanced probabilities observed in the DNS results.

While the qualitative comparisons between streamlines of the restricted Euler system and joint-PDFs from DNS are encouraging for both heavy and light particles, quantitative comparison of stationary statistics cannot be accomplished without models for the neglected unclosed terms \cite{Meneveau2011}. Besides the pressure Hessian and viscous Laplacian, additional modeling work is likely necessary for the additional terms introduced for inertial trajectories, namely $\partial\left[\left(v_i-u_i\right)A_{ij}\right]/\partial x_k$.

%\section{Analysis}

Many of the features of the inertial restricted Euler system  can be investigated analytically. A salient feature of the original restricted Euler equation ($\alpha = 0$) is the invariant $Q^3 + \tfrac{27}{4}R^2$ \cite{Vieillefosse1982,Cantwell1992}. For the extended system given by \eqref{eq:QR},
\begin{equation}
\frac{d}{dt}\left(Q^3 + \frac{27}{4}R^2\right) = - 2 \alpha Q \left( Q^3 + \frac{27}{4}R^2 \right),
\end{equation}
so that for the particular choice $Q^3 + \tfrac{27}{4}R^2 = 0$, this remains an invariant of the dynamics. In particular, this means that the so-called Vieillefosse tail, $Q_v(R) = - \left(\tfrac{27}{4}\right)^{1/3} R^{2/3}$, is an invariant manifold for all values of $\alpha$.

It is straightforward to show that \eqref{eq:QR} has two fixed points, one at the origin and another at $R_0 = -2/\alpha^3$, $Q_0 = -3/\alpha^2$.
The fixed point away from the origin lies on the Vieillefosse tail, i.e., $Q_0^3 + \tfrac{27}{4}R_0^2 = 0$. For heavy particles, the fixed point lies in the third quadrant on an $RQ$ graph ($R_0<0$, $Q_0<0$), while for light particles, the fixed point lies in the fourth quadrant ($R_0>0$, $Q_0<0$). Linear stability analysis of this fixed point reveals eigenvalues of $\lambda_1 = 6/\alpha$ and $\lambda_2 = 1/\alpha$ with (unnormalized) eigenvectors $\mathbf{e}^{(1)} = \left( 1, -\tfrac{3}{2}\alpha \right)^T$ and $\mathbf{e}^{(2)} = \left( 1, \alpha \right)^T$. The fixed point is unstable for heavy particles and stable for light particles. The slope of the Vieillefosse tail at the fixed point is $\left.dQ_v/dR\right|_{Q_0} = \alpha$, so that the eigenvector associated with the more weakly stable/unstable eigenvector points along the Vieillefosse manifold.

Along the Vieillefosse manifold, the dynamics are given by $dR/dt = \tfrac{3}{2^{1/3}} R^{4/3} + \tfrac{3}{2^{2/3}} \alpha R^{5/3}$, which for $\alpha \neq 0$ can be written as
\begin{equation}
\frac{dR}{dt} = \frac{6}{\alpha^4} \left[ \left(\frac{R}{R_0}\right)^{4/3} - \left(\frac{R}{R_0}\right)^{5/3} \right].
\end{equation}
This shows the reinforcement of finite-time singularity in the fourth quadrant for heavy particles and the introduction of another singularity in the third quadrant, for $R < R_0 < 0$. It also shows that the finite-time singularity along the Vieillefosse manifold is regularized for light particles due to the stable fixed point.

The linear stability of the Vieillefosse manifold is examined by considering the trajectory $Q(R) = Q_v(R) + \epsilon(R)$. Using $d\ln\epsilon/dt = d\ln\epsilon/dR ~ dR/dt$, the linearized behavior of $\epsilon$ can be shown to be
\begin{equation}
\frac{d\ln\epsilon}{dt} = \left(2^{4/3} - 1\right) \alpha R^{1/3} \left( R^{1/3} - \left( \frac{32}{(16^{1/3} - 1)^3 \alpha^3} \right)^{1/3} \right).
\end{equation}
When $d\ln\epsilon/dt > 0$, the Vieillefosse line is an unstable manifold. When $d\ln\epsilon/dt < 0$ it is a stable manifold. The stability of the manifold changes sign twice: once at the origin, and also at the point
\begin{equation}
\left( R_s, Q_s \right) = \left( \frac{32}{\left(16^{1/3}-1\right)^3\alpha^3}, -\frac{12 (2)^{2/3}}{(16^{1/3}-1)^2\alpha^2}\right).
\end{equation}

For $\alpha > 0$ (heavy particles), the following chart summarizes the stability of the Vieillefosse manifold,
\begin{center}
\begin{tabular}{c |c| c}
$R < 0$ &  $0< R < \frac{32}{\left(16^{1/3}-1\right)^3\alpha^3}$ & $R > \frac{32}{\left(16^{1/3}-1\right)^3\alpha^3}$ \\
\hline
unstable & stable & unstable \\
\end{tabular}.
\end{center}
Meanwhile, for $\alpha < 0$ (light particles), the stability can be summarized as,
\begin{center}
\begin{tabular}{c |c| c}
$R < \frac{32}{\left(16^{1/3}-1\right)^3\alpha^3}$ &  $\frac{32}{\left(16^{1/3}-1\right)^3\alpha^3}< R < 0$ & $R > 0$ \\
\hline
stable & unstable & stable \\
\end{tabular}.
\end{center}

In conclusion, the extension of the restricted Euler system for velocity gradients along inertial particle paths yields qualitative agreement with basic trends seen from DNS when projected onto the RQ plane. The trends observed follow directly from first principles, i.e from the ``self-stretching'' properties of the nonlinear term in the Navier-Stokes and particle transport equations whose effects are elucidated here by neglecting all of the ``non-local'' spatial flux terms.  For these reasons, the model can be a good starting point for developing more complete models for velocity gradients along inertial particle trajectories for applications such as preferential (fractal) concentration  \cite{Bec2003,Bec2006,Esmaily-Moghadam2015} of heavy and light anisotropic particles \cite{Parsa2012,Chevillard2013} and deformation of liquid droplets \cite{Biferale2014} or bubbles.

%In addition to the restricted Euler model considered in this paper, the pressure Hessian and viscous Laplacian terms must be re-introduced. Various approaches have been proposed for these terms when considering Lagrangian fluid (tracer) trajectories \cite{Meneveau2011}. Additional modeling efforts, however, are likely needed for the additional inertial term, $\partial\left[\left(v_i-u_i\right)A_{ij}\right]/\partial x_k$. The Gaussian closure approach \cite{Wilczek2014} provides one such possibility. 
%Low-dimensional representations of inertial particle dynamics in turbulence  with stationary statistics could be useful for studying the rate at which heavy and light particles preferentially concentrate as well as the fractal dimension to which they collapse \cite{Bec2003,Bec2006,Esmaily-Moghadam2015}. In addition, such a model could provide insights for how inertial corrections impact applications such as the deformation of immiscible droplets \cite{Biferale2014}.

%\section*{Acknowledgments}

The authors are very grateful to Luca Biferale and Federico Toschi for making their DNS simulation data available for the plots generated in figure 3. PJ was supported by a National Science Foundation Graduate Research Fellowship Program under Grant No.
DGE-1232825. CM's research was made possible by a grant from The Gulf of Mexico Research Initiative.

\bibliographystyle{apsrev4-1}
\bibliography{inertial_LaVelGrad.bib}

\end{document}